\documentclass[aps,prd,nofootinbib,amsmath,amssymb,showpacs,superscriptaddress,twocolumn]{revtex4}
\usepackage{txfonts}
\usepackage{graphicx}
\usepackage{dcolumn}
\usepackage{bm}
\usepackage{amssymb}
\usepackage{latexsym}
\usepackage[colorlinks, linkcolor=blue, citecolor=blue, urlcolor=blue]{hyperref}

\newcommand{\be}{\begin{equation}}
\newcommand{\ee}{\end{equation}}

\def\bea{\begin{eqnarray}}
\def\eea{\end{eqnarray}}
\begin{document}

\title{Holographic $\Lambda(t)$CDM model in a non-flat universe}

\author{Jing-Fei Zhang}
\affiliation{Department of Physics, College of Sciences,
Northeastern University, Shenyang 110004, China}
\author{Yang-Yang Li}
\affiliation{Department of Physics, College of Sciences,
Northeastern University, Shenyang 110004, China}
\author{Ying Liu}
\affiliation{Department of Physics, College of Sciences,
Northeastern University, Shenyang 110004, China}
\author{Sheng Zou}
\affiliation{Department of Physics, College of Sciences,
Northeastern University, Shenyang 110004, China}
\author{Xin Zhang\footnote{Corresponding author}}
\email{zhangxin@mail.neu.edu.cn} \affiliation{Department of Physics,
College of Sciences, Northeastern University, Shenyang 110004,
China} \affiliation{Center for High Energy Physics, Peking
University, Beijing 100080, China}

\begin{abstract}
The holographic $\Lambda(t)$CDM model in a non-flat universe is
studied in this paper. In this model, to keep the form of the
stress-energy of the vacuum required by general covariance, the
holographic vacuum is enforced to exchange energy with dark matter.
It is demonstrated that for the holographic model the best choice
for the IR cutoff of the effective quantum field theory is the event
horizon size of the universe. We derive the evolution equations of
the holographic $\Lambda(t)$CDM model in a non-flat universe. We
constrain the model by using the current observational data,
including the 557 Union2 type Ia supernovae data, the cosmic
microwave background anisotropy data from the 7-yr WMAP, and the
baryon acoustic oscillation data from the SDSS. Our fit results show
that the holographic $\Lambda(t)$CDM model tends to favor a
spatially closed universe (the best-fit value of $\Omega_{k0}$ is
$-0.042$), and the $95\%$ confidence level range for the spatial
curvature is $-0.101<\Omega_{k0}<0.040$. We show that the
interaction between the holographic vacuum and dark matter induces
an energy flow of which the direction is first from vacuum to dark
matter and then from dark matter to vacuum. Thus, the holographic
$\Lambda(t)$CDM model is just a time-varying vacuum energy scenario
in which the interaction between vacuum and dark matter changes sign
during the expansion of the universe.
\end{abstract}



\maketitle

\section{Introduction}\label{sec:Intro}

Dark energy has been one of the most important themes in modern
physics. However, today, we are still far from thoroughly
understanding the nature of dark energy~\cite{DErev}. The
cosmological constant $\Lambda$, posited in 1917 and later rejected
by Einstein~\cite{Einstein17}, is an important candidate for dark
energy, because it can provide a nice explanation for the
accelerating universe and can fit the observational data
well~\cite{wmap7}. The cosmological model containing the
cosmological constant $\Lambda$ and cold dark matter (CDM) is known
as the $\Lambda$CDM model, which is viewed as the most important
cosmological model today in the cosmology community. Nevertheless,
the cosmological constant is suffering from severe theoretical
challenge: one cannot understand why the theoretical value of
$\Lambda$ from the current framework of physics is greater than the
observational value by many orders of magnitude~\cite{ccrev}. It is
known that the cosmological constant is equivalent to the vacuum
energy, and so its value is determined by the sum of the zero-point
energy of each mode of all the quantum fields; thus, we have
$\rho_\Lambda\simeq k_{\rm max}^4/(16\pi^2)$, where $k_{\rm max}$ is
the imposed momentum ultraviolet (UV) cutoff. Taking the UV cutoff
to be the Planck scale ($\approx 10^{19}$ GeV), where one expects
quantum field theory in a classical spacetime metric to breakdown,
the vacuum energy density would exceed the critical density by some
120 orders of magnitude.

Obviously, the key to the problem is gravity. One should not have
calculated the value of $\Lambda$ or the vacuum energy density in a
context without gravity. Actually, it is conjectured that the
cosmological constant problem would be solved when a full theory of
quantum gravity is established. However, in the present day that we
have no a quantum gravity theory, how could we understand the
cosmological constant problem from a point of view of quantum
gravity? In fact, this attempt has begun---a typical example is the
holographic dark energy model~\cite{Li04} which originates from the
holographic principle~\cite{holop} of quantum gravity. It is
expected that the theoretical and phenomenological studies on
holographic dark energy might provide significant clues for the
bottom-up exploration of a full quantum theory of gravity.

When considering gravity in a quantum field system, the conventional
local quantum field theory would break down due to the too many
degrees of freedom that would cause the formation of a black hole.
However, once the holographic principle is considered, the number of
degrees of freedom could be reduced. One could put an energy bound
on the vacuum energy density, $\rho_\Lambda L^3\leq M_{\rm Pl}^2 L$,
where $M_{\rm Pl}$ is the reduced Planck mass, which implies that
the total energy in a spatial region with size $L$ should not exceed
the mass of a black hole with the same size~\cite{Cohen99}. The
largest length size compatible with this bound is the infrared (IR)
cutoff size of this effective quantum field theory. Evidently, the
holographic principle gives rise to a dark energy model basing on
the effective quantum field theory with a UV/IR duality. From the
UV/IR correspondence, the UV problem of dark energy can be converted
into an IR problem. A given IR scale can saturate that bound, and so
one can write the dark energy density as $\rho_\Lambda=3c^2M_{\rm
Pl}^2L^{-2}$~\cite{Li04}, where $c$ is a phenomenological parameter
(dimensionless) characterizing all of the uncertainties of the
theory. This indicates that the UV cutoff of the theory would not be
fixed but run with the evolution of the IR cutoff, i.e., $k_{\rm
max}\propto L^{-1/2}$. The original holographic dark energy model
chooses the event horizon of the universe as the IR cutoff of the
theory, explaining the fine-tuning problem and the coincidence
problem at the same time in some degree~\cite{Li04}. Actually, it is
clear to see that the holographic dark energy is essentially a
holographic vacuum energy~\cite{Zhang:2006av}. However, this
holographic vacuum energy does not behave like a usual vacuum
energy, owing to the fact that its equation of state (EOS) parameter
$w$ is not equal to $-1$. To keep the form of the stress-energy of
the vacuum, $T_\Lambda^{\mu\nu}=\rho_\Lambda g^{\mu\nu}$, required
by general covariance, one possible way is to let the vacuum
exchange energy with dark matter. This requires that we fix the EOS,
$w=-1$, for the holographic vacuum energy, resulting in the
continuity equations for $\Lambda$ and CDM, $\dot{\rho}_\Lambda=-Q$
and $\dot{\rho}_{\rm m}+3H\rho_{\rm m}=Q$, where $Q$ describes the
interaction between $\Lambda$ and CDM; note that, although
$\rho_{\rm m}$ includes densities of cold dark matter and baryon
matter, in this place we use $\rho_{\rm m}$ to approximately
describe dark matter density due to the fact that the density of
baryon matter is much less than that of dark matter. We call this
model the ``holographic $\Lambda(t)$CDM model.''

In fact, the possibility that $\Lambda$ is not a real constant but
is time variable or dynamical was considered many years
ago~\cite{varycc}. Usually, one specifies a time-dependence form for
$\Lambda(t)$ by hand and then establishes a phenomenological
$\Lambda(t)$CDM model~\cite{varycc2}. In our holographic
$\Lambda(t)$CDM model, however, $\Lambda(t)$ originates from the
holographic principle of quantum gravity, and the scenario is
established based on the effective quantum field theory combining
with the fundamental principle of quantum gravity. Now, as usual,
the problem becomes how to choose an appropriate IR cutoff for the
theory. As mentioned above, the original holographic dark energy
model takes the event horizon as the IR cutoff, and this model
setting obtained great successes in both theoretical and
observational aspects~\cite{holoext,holofit}. Nevertheless, a
criticism concerning the causality problem arose for this model: the
existence of the event horizon should be a consequence but should
not be a premise of a dark energy model. This criticism gave rise to
other versions of holographic dark energy, such as agegraphic dark
energy model~\cite{ADE} and Ricci dark energy model~\cite{RDE}.
However, a recent study on inflation and quantum
mechanics~\cite{Huang:2012eu} may remove the causality problem from
the holographic dark energy model. In Ref.~\cite{Huang:2012eu}, the
authors showed that, if inflation indeed happened in the early
times, the quantum no-cloning theorem requires that the existence of
event horizon is a must. So, once inflation is accommodated in a
cosmological scenario, the existence of the event horizon could be
adopted as a premise of a holographic dark energy model. On the
other hand, the holographic dark energy model based on the event
horizon is much better than other versions of the holographic model
in fitting the observational data~\cite{holocompare}. For the
various versions of the holographic $\Lambda(t)$CDM model (i.e., the
versions taking the IR cutoff as the Hubble scale, particle horizon,
event horizon, age of the universe, conformal time of the universe,
etc), it has also been shown that the version concerning the event
horizon fits the observational data best~\cite{Chen:2011rz}. Thus,
based on these facts, in this paper, we only consider the
holographic $\Lambda(t)$CDM model in which the event horizon of the
universe provides the effective quantum field theory with the IR
length-scale cutoff.

In this paper, we study the holographic $\Lambda(t)$CDM model in a
non-flat universe. It is well known that the flatness of the
observable universe is one of the predictions of conventional
inflation models. Though inflation theoretically produces
$\Omega_{k0}$ on the order of the magnitude of quantum fluctuations,
namely, $\Omega_{k0}\sim 10^{-5}$, the current observational limit
on $\Omega_{k0}$ is of order $10^{-2}$~\cite{wmap7}. In addition,
since the spatial curvature is degenerate with the parameters of
dark energy models, there is a revived and growing interest in
studying dark energy models with spatial curvature. In the
following, we shall first derive the evolution equations of the
holographic $\Lambda(t)$CDM model and then place the current
observational constraints on the model.

The paper is organized as follows: In Sec.~\ref{sec:model} we will
derive the evolution equations of the holographic $\Lambda(t)$CDM
model in a non-flat universe, the cosmological constraints and the
result discussions are presented in Sec.~\ref{sec:obs}, and the
conclusion is given in Sec.~\ref{sec:concl}.

\section{The holographic $\Lambda(t)$CDM model}\label{sec:model}

Consider the Friedmann-Robertson-Walker universe with the metric
\begin{equation}
ds^2=-dt^2+a(t)^2\left({dr^2\over 1-kr^2}+r^2
d\theta^2+r^2\sin^2\theta d\phi^2\right),
\end{equation}
where $k=1$, 0, $-1$ for closed, flat, and open geometries,
respectively, and $a(t)$ is the scale factor of the universe with
the convention $a(t_0)=1$. The Friedmann equation is
\begin{equation}
H^2={8\pi G\over 3}(\rho_{\rm m}+\rho_\Lambda) -{k\over
a^2},\label{Feq}
\end{equation}
where $H=\dot{a}/a$ is the Hubble parameter. Since we focus only on
the late-time evolution of the universe, the radiation component
$\rho_{\rm rad}$ is negligible.

By definition of the holographic vacuum energy, we have
\begin{equation}
\rho_\Lambda=3c^2 M_{\rm Pl}^2 L^{-2},
\end{equation}
where $L$ is the IR length-scale cutoff of the theory,
\begin{equation}
L=ar(t),
\end{equation}
with $r(t)$ determined by
\begin{equation}
\int_0^{r(t)} {dr \over \sqrt{1-kr^2}}=\int_t^{+\infty}{dt\over
a(t)}.
\end{equation}
From the above equation, one can easily obtain
\begin{equation}
r(t)={1\over\sqrt{k}}\sin \Big(\sqrt{k}\int_t^{+\infty} {dt \over
a}\Big)={1\over\sqrt{k}}\sin \Big(\sqrt{k}\int_{a(t)}^{+\infty} {da
\over {Ha^2}}\Big).\label{r1}
\end{equation}

The Friedmann equation (\ref{Feq}) can be rewritten as
\begin{equation}
\Omega_{\rm m}+\Omega_\Lambda+\Omega_k=1,
\end{equation}
where
\begin{equation}
\Omega_k={-k\over H^2 a^2}={\Omega_{k0}a^{-2}\over (H/H_0)^2},
\end{equation}
and
\begin{equation}
\Omega_\Lambda={\rho_\Lambda\over 3M_{\rm Pl}^2 H^2}={c^2\over H^2
L^2}.\label{OmegaL}
\end{equation}
From Eq.~(\ref{OmegaL}), one has $L=c/(H\sqrt{\Omega_\Lambda})$, and
thus
\begin{equation}
r(t)={L\over a}={c\over aH\sqrt{\Omega_\Lambda}}.\label{r2}
\end{equation}
Combining Eqs.~(\ref{r1}) and (\ref{r2}), we have
\begin{equation}
\arcsin{c\sqrt{k}\over
aH\sqrt{\Omega_{\Lambda}}}=\sqrt{k}\int_t^{+\infty}{dt\over a}.
\end{equation}
Taking the derivatives with respect to time $t$ for the both sides
of the above equation, we obtain
\begin{equation}
{\dot\Omega_{\Lambda}\over2\Omega_{\Lambda}H}+1+{\dot{H}\over
H^2}=\sqrt{{\Omega_{\Lambda}\over c^2}+\Omega_k}.\label{deqa}
\end{equation}

In our model, as mentioned above, the vacuum exchanges energy with
dark matter, so the continuity equations for them are
\begin{equation}
\dot{\rho}_\Lambda=-Q,\label{coneq1}
\end{equation}
\begin{equation}
\dot\rho_{\rm m}+3H\rho_{\rm m}=Q.\label{coneq2}
\end{equation}
This means that $\dot\rho_{\Lambda}+\dot\rho_{\rm m}+3H\rho_{\rm
m}=0$ or $\dot\rho_{\rm cr}-\dot\rho_k+3H(\rho_{\rm
cr}-\rho_{\Lambda}-\rho_k)=0$, where $\rho_{\rm cr}=3M_{\rm
Pl}^2H^2$ is the critical density of the universe. Using
$\dot\rho_{\rm cr}=2(\dot H/H)\rho_{\rm cr}$ and
$\dot\rho_k=-2H\rho_k$, one gets
\begin{equation}
{\dot H\over H^2}={1\over
2}(3\Omega_\Lambda+\Omega_k-3).\label{deqb1}
\end{equation}
Substituting this relation into Eq.~(\ref{deqa}), one obtains the
expression
\begin{equation}
\dot\Omega_\Lambda=2H\Omega_\Lambda\left[\sqrt{{\Omega_\Lambda\over
c^2}+\Omega_k}-{1\over
2}(3\Omega_\Lambda+\Omega_k-1)\right].\label{deqb2}
\end{equation}
Differential equations (\ref{deqb1}) and (\ref{deqb2}) govern the
cosmological evolution of the holographic $\Lambda(t)$CDM model. In
practice, we replace the variable $t$ with the variable $z$, and
derive the following two differential equations
\begin{equation}
{1\over (H/H_0)}{d(H/H_0)\over dz}=-{1\over
2(1+z)}(3\Omega_\Lambda+\Omega_k-3),\label{deq1}
\end{equation}
\begin{equation}
{d\Omega_\Lambda\over dz}=-{\Omega_\Lambda\over
1+z}\left(2\sqrt{{\Omega_\Lambda\over
c^2}+\Omega_k}-3\Omega_\Lambda-\Omega_k+1\right).\label{deq2}
\end{equation}
Solving the differential equations (\ref{deq1}) and (\ref{deq2}), we
can learn the evolution behavior of $H(z)$ and $\Omega_\Lambda(z)$,
and then various observables can also be obtained. We will
numerically solve these two equations and use the solutions to fit
with the observational data.

One may also be interested in the effective EOS parameters of the
holographic vacuum energy and the cold dark matter. Equations
(\ref{coneq1}) and (\ref{coneq2}) can be rewritten as
\begin{equation}
\dot{\rho}_{\Lambda}+3H(1+w_{\Lambda}^{\rm eff})\rho_{\Lambda}=0,
\end{equation}
\begin{equation}
\dot{\rho}_{\rm m}+3H(1+w_{\rm m}^{\rm eff})\rho_{\rm m}=0,
\end{equation}
where $w_{\Lambda}^{\rm eff}=-1+Q/(3H\rho_\Lambda)$ and $w_{\rm
m}^{\rm eff}=-Q/(3H\rho_{\rm m})$ are the effective EOS parameters
for holographic vacuum energy and cold dark matter, respectively.
One can easily obtain the expressions
\begin{equation}
w_{\Lambda}^{\rm eff}=-{1\over 3}\left(2\sqrt{{\Omega_\Lambda\over
c^2}+\Omega_k}+1\right),\label{eos1}
\end{equation}
\begin{equation}
w_{\rm m}^{\rm eff}={2\Omega_\Lambda\over
3(1-\Omega_\Lambda-\Omega_k)}\left(\sqrt{{\Omega_\Lambda\over
c^2}+\Omega_k}-1\right).\label{eos2}
\end{equation}
The deceleration parameter $q\equiv-\ddot{a}/(aH^2)$ can also be
easily derived,
\begin{equation}
q=-{1\over 2}(3\Omega_\Lambda+\Omega_k-1).
\end{equation}
So, one sees that, once the solutions $H(z)$ and $\Omega_\Lambda(z)$
are obtained from the differential equations (\ref{deq1}) and
(\ref{deq2}), these quantities are known to us.

\section{Cosmological constraints}\label{sec:obs}

In this section, we will constrain the holographic $\Lambda(t)$CDM
model by using the current observational data. We will use the 557
SN data from the Union2 dataset, the CMB data from the WMAP 7-year
observation, and the BAO data from the SDSS. We will obtain the
best-fitted parameters and likelihoods by using a Markov Chain Monte
Carlo (MCMC) method.

We use the data points of the 557 Union2 SN compiled in
Ref.~\cite{14Amanullah:2010vv}. The theoretical distance modulus is
defined as
\begin{equation}
\label{eq11} \mu_{\rm th}(z_i)\equiv5\log_{10} D_L(z_i)+\mu_0,
\end{equation}
where $\mu_0\equiv42.38-5\log_{10} h$ with $h$ the Hubble constant
$H_0$ in units of 100 km s$^{-1}$ Mpc$^{-1}$, and the Hubble-free
luminosity distance
\begin{equation}
D_L(z)={1+z\over \sqrt{|\Omega_{k0}|}}\textrm{sinn}\left(
\sqrt{|\Omega_{k0}|}\int_0^z{dz'\over E(z')} \right),
\end{equation}
where $E(z)\equiv H(z)/H_0$, and
\begin{displaymath}
{\textrm{sinn}\left(\sqrt{|\Omega_{k0}|}x\right)\over
\sqrt{|\Omega_{k0}|}} = \left\{
\begin{array}{ll}
{\textrm{sin}(\sqrt{|\Omega_{k0}|}x)/
\sqrt{|\Omega_{k0}|}}, & \textrm{if $\Omega_{k0}<0$},\\
x, & \textrm{if $\Omega_{k0}=0$},\\
{\textrm{sinh}(\sqrt{|\Omega_{k0}|}x)/
 \sqrt{|\Omega_{k0}|}}, &
\textrm{if $\Omega_{k0}>0$}.
\end{array} \right.
\end{displaymath}
The $\chi^2$ function for the 557 Union2 SN data is given by
\begin{equation}
\label{eq13} \chi^2_{\rm
SN}=\sum\limits_{i=1}^{557}\frac{\left[\mu_{\rm obs}(z_i)-\mu_{\rm
th}(z_i)\right]^2}{\sigma^2(z_i)},
\end{equation}
where $\sigma$ is the corresponding $1\sigma$ error of distance
modulus for each supernova. The parameter $\mu_0$ is a nuisance
parameter and one can expand Eq.~(\ref{eq13}) as
\begin{equation}
\label{eq14} \chi^2_{\rm SN}=A-2\mu_0 B+\mu_0^2 C,
\end{equation}
where $A$, $B$ and $C$ are defined as in
Ref.~\cite{Nesseris:2005ur}.
Evidently, Eq.~(\ref{eq14}) has a minimum for $\mu_0=B/C$ at
\begin{equation}
\label{eq15} \tilde{\chi}^2_{\rm SN}=A-\frac{B^2}{C}.
\end{equation}
Since $\chi^2_{\rm SN,\,min}=\tilde{\chi}^2_{\rm SN,\,min}$, instead
minimizing $\chi_{\rm SN}^2$ we will minimize $\tilde{\chi}^2_{\rm
SN}$ which is independent of the nuisance parameter $\mu_0$.

Next, we consider the cosmological observational data from WMAP and
SDSS. For the WMAP data, we use the CMB shift parameter $R$; for the
SDSS data, we use the parameter $A$ of the BAO measurement. It is
widely believed that both $R$ and $A$ are nearly model-independent
and contain essential information of the full WMAP CMB and SDSS BAO
data \cite{55Wang:2006ts}. The shift parameter $R$ is given by
\cite{55Wang:2006ts,54Bond:1997wr}
 \begin{equation}\label{eq31}
R\equiv{\sqrt{\Omega_{\rm m0}}\over
\sqrt{|\Omega_{k0}|}}\textrm{sinn}\left(\sqrt{|\Omega_{k0}|}\int_0^{z_\ast}{dz'\over
E(z')} \right),
 \end{equation}
where the redshift of recombination $z_\ast=1091.3$, from the WMAP
7-year data \cite{wmap7}. The shift parameter $R$ relates the
angular diameter distance to the last scattering surface, the
comoving size of the sound horizon at $z_\ast$ and the angular scale
of the first acoustic peak in the CMB power spectrum of temperature
fluctuations \cite{55Wang:2006ts,54Bond:1997wr}. The value of $R$ is
$1.725\pm0.018$, from the WMAP 7-year data \cite{wmap7}. The
distance parameter $A$ from the measurement of the BAO peak in the
distribution of SDSS luminous red galaxies \cite{57Tegmark:2003ud}
is given by
 \begin{equation}\label{eq32}
   A\equiv{\sqrt{\Omega_{\rm m0}}\over
{E(z_{\rm b})}^{1\over3}}\left[{1\over z_{\rm
b}\sqrt{|\Omega_{k0}|}}\textrm{sinn}\left(\sqrt{|\Omega_{k0}|}\int_0^{z_{\rm
b}}{dz'\over E(z')} \right)\right]^{2\over3},
 \end{equation}
where $z_{\rm b}=0.35$. In Ref.~\cite{58Eisenstein:2005su}, the
value of $A$ has been determined to be $0.469\ (n_{\rm
s}/0.98)^{-0.35}\pm0.017$. Here, the scalar spectral index $n_{\rm
s}$ is taken to be $0.963$, from the WMAP 7-year data \cite{wmap7}.
So the total $\chi^2$ is given by
 \begin{equation}\label{eq33}
   \chi^2=\tilde{\chi}_{\rm SN}^2+\chi_{\rm CMB}^2+\chi_{\rm BAO}^2,
 \end{equation}
where $\tilde{\chi}_{\rm SN}^2$ is given by (\ref{eq15}), $\chi_{\rm
CMB}^2=(R-R_{\rm obs})^2/\sigma_R^2$ and $\chi_{\rm BAO}^2=(A-A_{\rm
obs})^2/\sigma_A^2$. The best-fitted model parameters are determined
by minimizing the total $\chi^2$.

\begin{table*} \caption{The fit results of the $\Lambda$CDM and holographic $\Lambda(t)$CDM
models.}\label{table1}
\begin{center}
\label{table1}
\begin{tabular}{ccccccc}
  \hline\hline
Model                &           $c$                   &             $\Omega_{\rm m0}$      &          $\Omega_{k0}$        &    $\chi^2_{\rm min}$   \\
  \hline
  flat~$\Lambda$CDM               &    N/A                           &~~ $0.270^{+0.014}_{-0.013}$ ~~ &          N/A                       &~~~    542.919    ~~~\\
  \hline
 non-flat~$\Lambda$CDM                &              N/A                 &~~ $0.274^{+0.025}_{-0.024}$ ~~ & ~~ $-0.004^{+0.012}_{-0.012}$ ~~& ~~~   542.699    ~~~\\
  \hline
  flat~$\Lambda(t)$CDM         &~~ $0.694^{+0.030}_{-0.032}$ ~~&~~ $0.256^{+0.023}_{-0.022}$~~ & ~~             N/A               ~~& ~~~   545.080    ~~~\\
  \hline
  non-flat~$\Lambda(t)$CDM     &~~ $0.766^{+0.103}_{-0.097}$ ~~&~~ $0.275^{+0.039}_{-0.036}$~~ & ~~ $-0.042^{+0.053}_{-0.044}$ ~~ & ~~~   542.803    ~~~\\
    \hline\hline
\end{tabular}
\end{center}
\end{table*}

\begin{figure*}[htbp]
\centering
\begin{center}
$\begin{array}{c@{\hspace{0.2in}}c} \multicolumn{1}{l}{\mbox{}} &
\multicolumn{1}{l}{\mbox{}} \\
\includegraphics[scale=0.25]{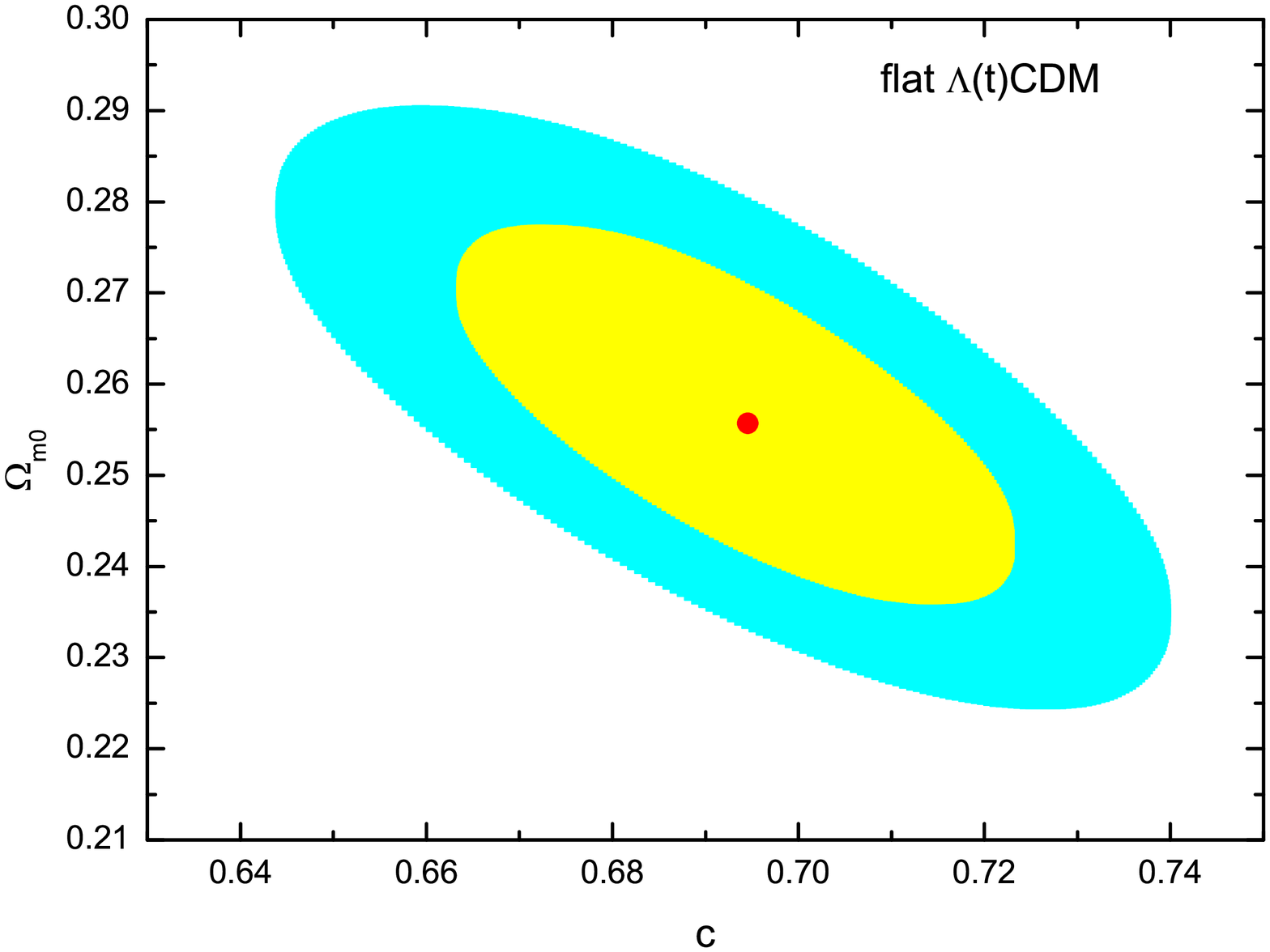} &\includegraphics[scale=0.25]{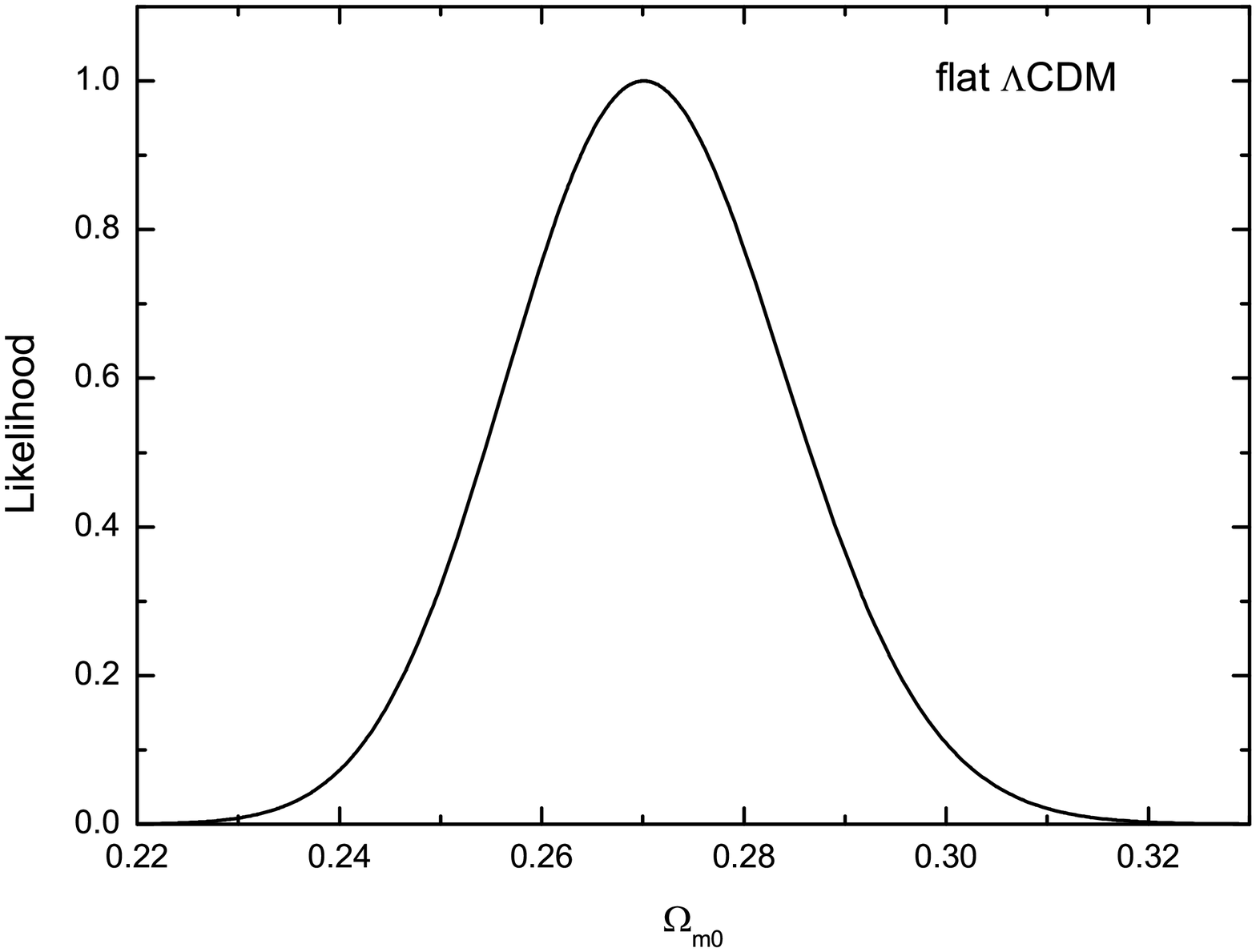} \\
\end{array}$
\end{center}
\caption[]{\small \label{fig1}The observational constraints on the
holographic $\Lambda(t)$CDM model and the $\Lambda$CDM model in a
flat universe. Left: the probability contours at $68\%$ and $95\%$
CLs in the $c$--$\Omega_{\rm m0}$ plane for the holographic
$\Lambda(t)$CDM model. Right: The one-dimensional likelihood
distribution of the parameter $\Omega_{\rm m0}$ for the $\Lambda$CDM
model.}
\end{figure*}

\begin{figure*}[htbp]
\centering
\begin{center}
$\begin{array}{c@{\hspace{0.2in}}c} \multicolumn{1}{l}{\mbox{}} &
\multicolumn{1}{l}{\mbox{}} \\
\includegraphics[scale=0.25]{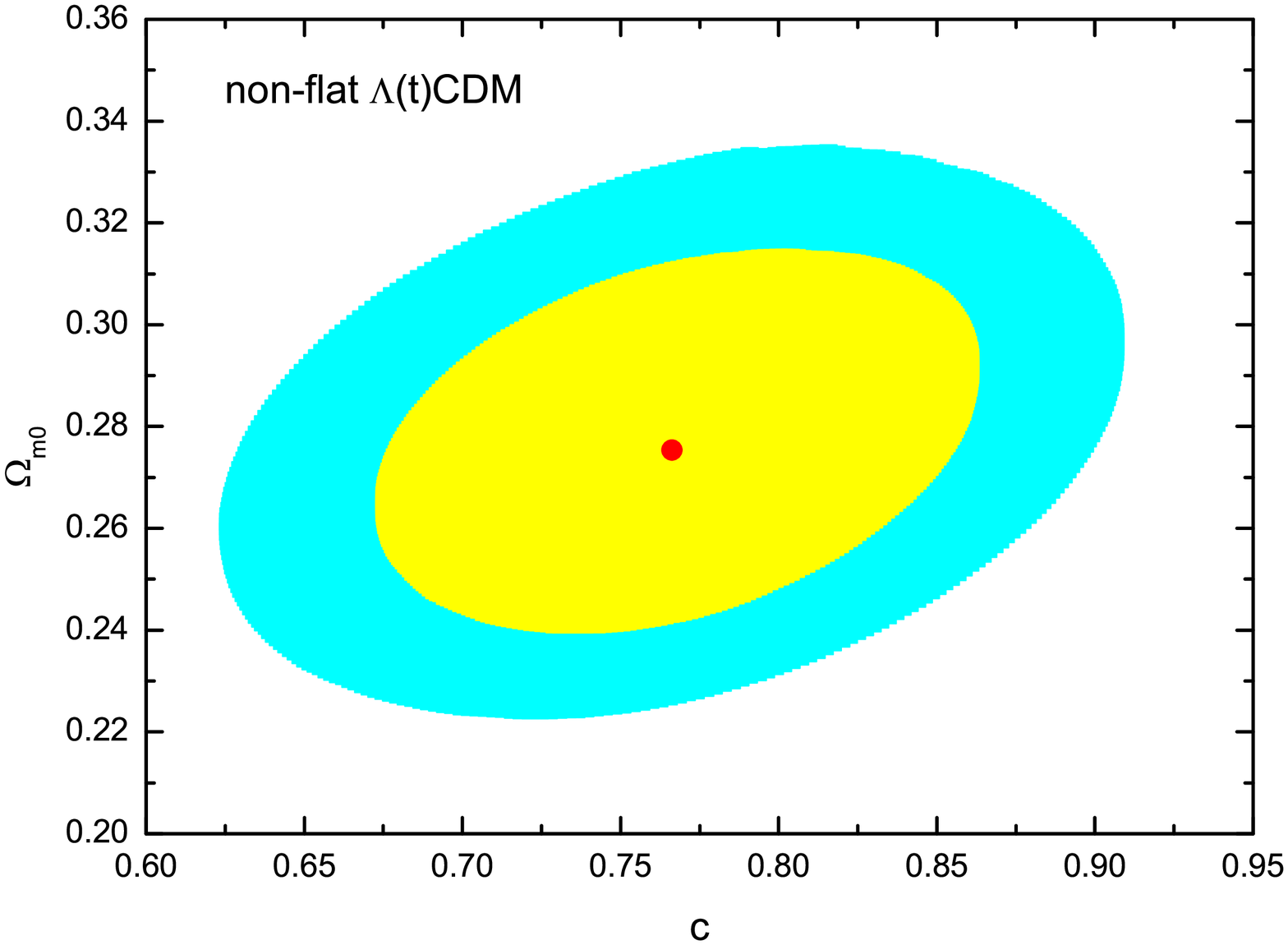} &\includegraphics[scale=0.25]{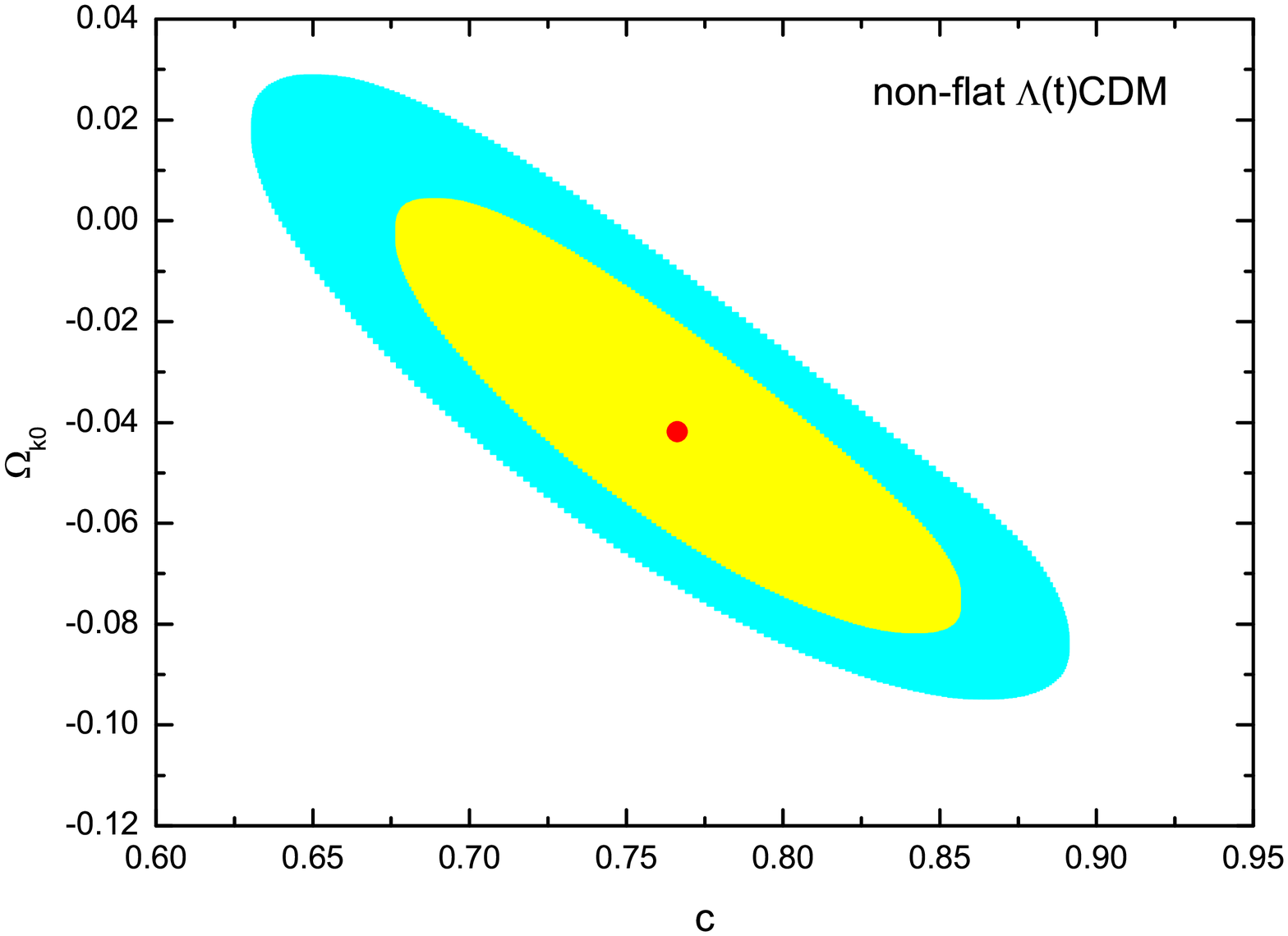} \\
\end{array}$
$\begin{array}{c@{\hspace{0.2in}}c} \multicolumn{1}{l}{\mbox{}} &
\multicolumn{1}{l}{\mbox{}} \\
\includegraphics[scale=0.25]{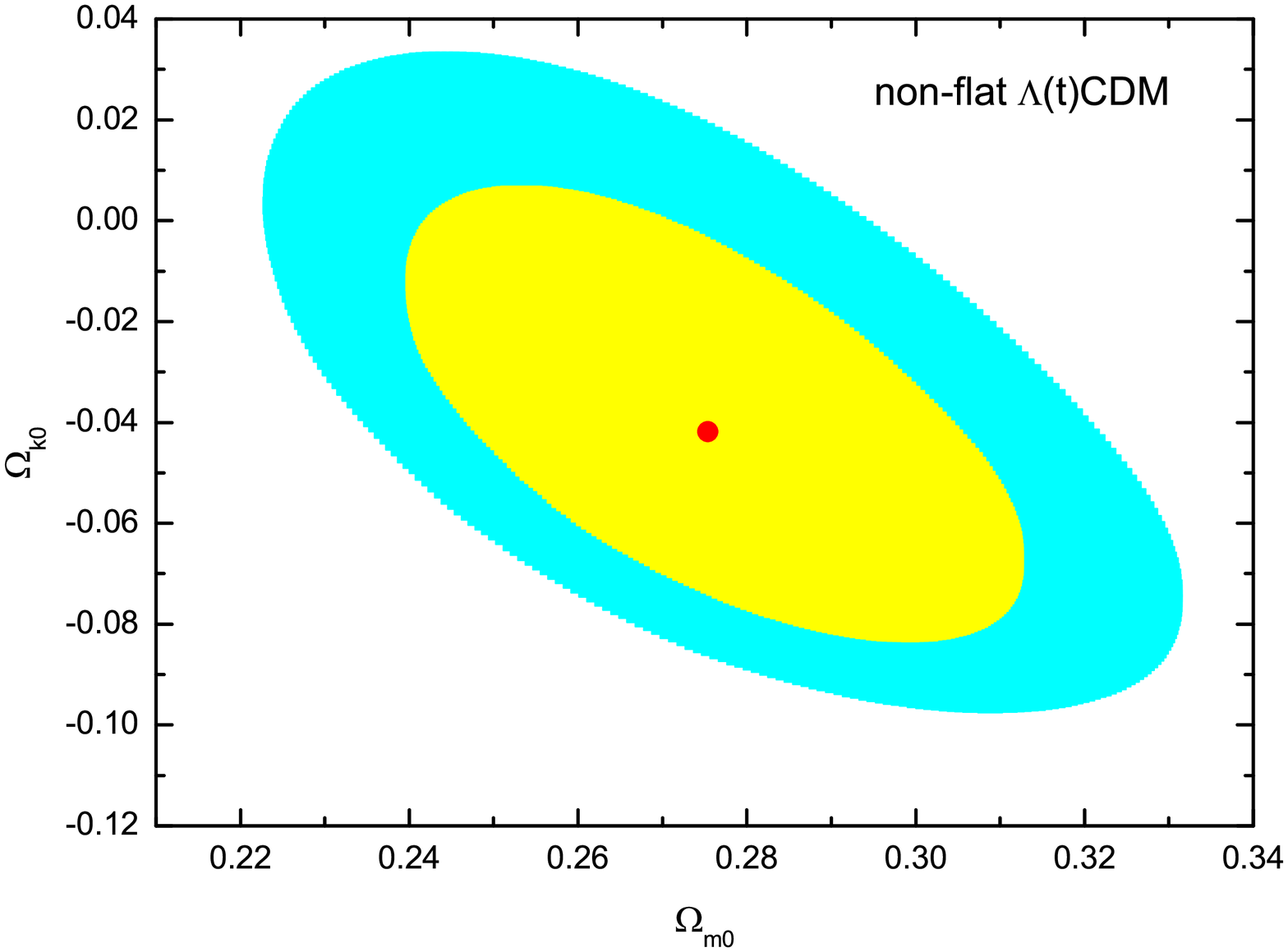} &\includegraphics[scale=0.25]{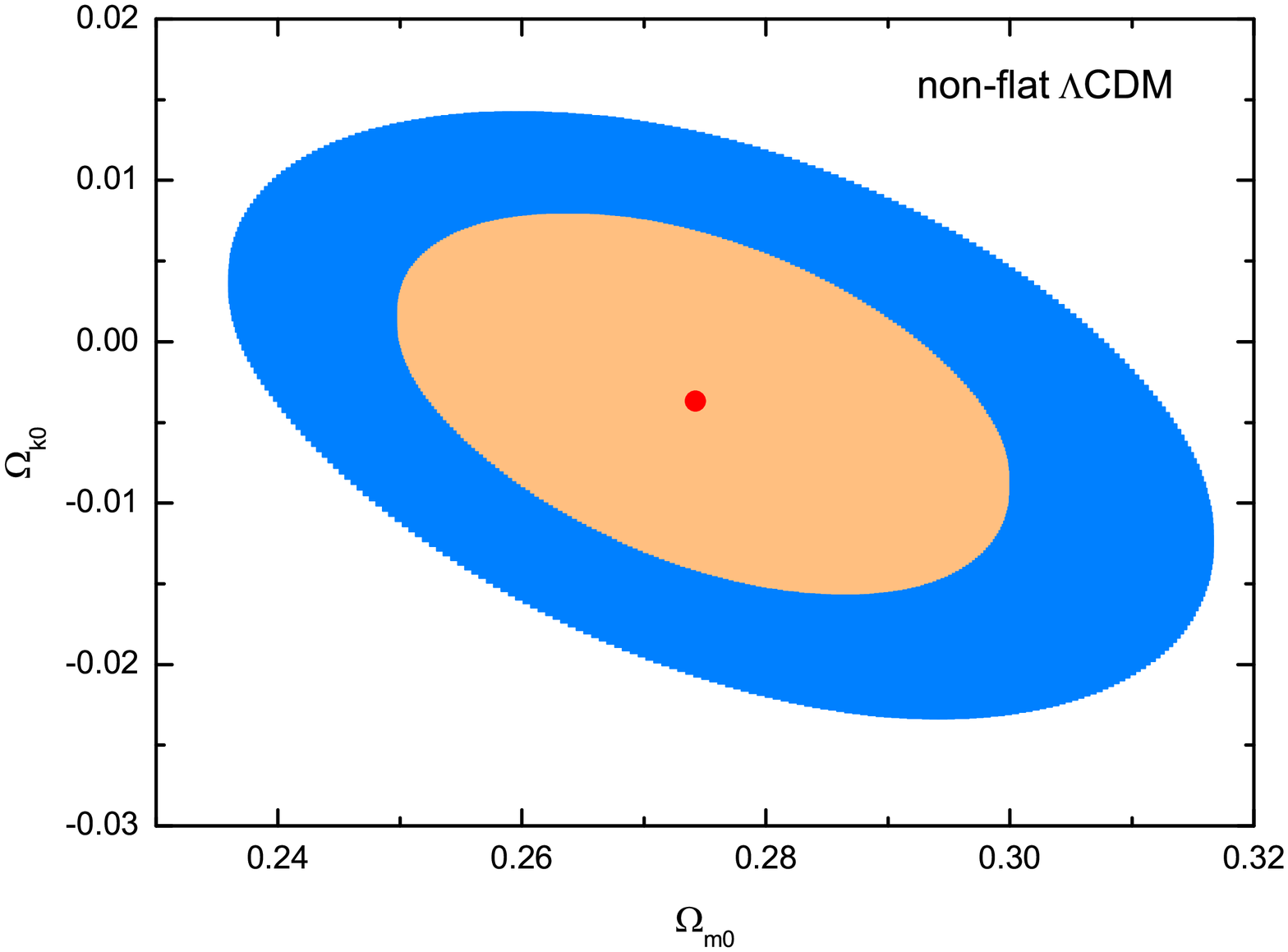} \\
\end{array}$
\end{center}
\caption[]{\small \label{fig2}The observational constraints on the
holographic $\Lambda(t)$CDM model and the $\Lambda$CDM model in a
non-flat universe. Upper-left, upper-right and lower-left: the
probability contours at $68\%$ and $95\%$ CLs in the
$c$--$\Omega_{\rm m0}$, $c$--$\Omega_{k0}$ and $\Omega_{\rm
m0}$--$\Omega_{k0}$ planes for the holographic $\Lambda(t)$CDM
model. Lower-right: the probability contours at $68\%$ and $95\%$
CLs in the $\Omega_{\rm m0}$--$\Omega_{k0}$ plane for the
$\Lambda$CDM model.}
\end{figure*}

Now, we fit the holographic $\Lambda(t)$CDM model to the
observational data. We use the MCMC method and finally we obtain the
best-fit values and 1$\sigma$, 2$\sigma$, and 3$\sigma$ values for
the model parameters. Since the $\Lambda$CDM model is an important
reference model for the studies of dark energy models, we also fit
the standard $\Lambda$CDM model to the same data for comparison. The
models are studied in the cases of flat and non-flat universes,
respectively. So, the calculations are performed for four
cases---flat $\Lambda$CDM, non-flat $\Lambda$CDM, flat
$\Lambda(t)$CDM, and non-flat $\Lambda(t)$CDM. The best-fit and
1$\sigma$ values of the parameters with $\chi^2_{\rm min}$ of the
four models are all presented in Table~\ref{table1}.

For the flat holographic $\Lambda(t)$CDM model, we obtain the
best-fit values of the parameters: $c=0.694$ and $\Omega_{\rm
m0}=0.256$, and the corresponding minimal $\chi^2$ is $\chi^2_{\rm
min}=545.080$. Obviously, the $\chi^2_{\rm min}$ value of this model
is much greater than that of the flat $\Lambda$CDM model, even
though in the case of number of parameters being greater by one. The
fit value of $\Omega_{\rm m0}$ is also a little bit small, 0.256,
evidently less than that of the $\Lambda$CDM model, 0.270. Now, let
us see the fit results of the holographic $\Lambda(t)$CDM model in
the case of a non-flat universe. For this case, we have $c=0.766$,
$\Omega_{\rm m0}=0.275$, $\Omega_{k0}=-0.042$, and $\chi^2_{\rm
min}=542.803$. Comparing to the non-flat $\Lambda$CDM model, we find
that the $\chi^2_{\rm min}$ values are similar (for the $\Lambda$CDM
model, $\chi^2_{\rm min}=542.699$), and the $\Omega_{\rm m0}$
best-fit values are also very similar (for the $\Lambda$CDM model,
$\Omega_{\rm m0}=0.274$). Comparing the holographic $\Lambda(t)$CDM
model in the flat and non-flat universes, we find that the spatial
curvature plays an important role in the model in fitting the data.
Once adding the parameter $\Omega_{k0}$ in the model, the
$\chi^2_{\rm min}$ value decreases by a very distinct amount, and
correspondingly, of course, the ranges of the parameters amplify to
some extent as usual. We find that the holographic $\Lambda(t)$CDM
model tends to favor a spatially closed universe (the best-fit value
of $\Omega_{k0}$ is $-0.042$), and we find that the $95\%$
confidence level (CL) range for the spatial curvature is
$-0.101<\Omega_{k0}<0.040$.

Figures \ref{fig1} and \ref{fig2} show the probability contours at
$68\%$ and $95\%$ CLs in the parameter planes for the holographic
$\Lambda(t)$CDM model and the $\Lambda$CDM model in flat and
non-flat universes. In Fig.~\ref{fig1}, we show the constraint
results for the models in a flat universe. In the left panel we plot
the contours in the $c$--$\Omega_{\rm m0}$ plane for the
$\Lambda(t)$CDM model, and in the right panel we plot the
one-dimensional likelihood of the parameter $\Omega_{\rm m0}$ for
the $\Lambda$CDM model. We find that, for the flat $\Lambda(t)$CDM
model, the parameters $c$ and $\Omega_{\rm m0}$ are anti-correlated.
In Fig.~\ref{fig2}, we show the constraint results for the models in
a non-flat universe. In this figure, the contours in the
$c$--$\Omega_{\rm m0}$, $c$--$\Omega_{k0}$ and $\Omega_{\rm
m0}$--$\Omega_{k0}$ planes for the non-flat $\Lambda(t)$CDM model
are shown in the upper-left, upper-right and lower-left panels,
respectively, and the contours in the $\Omega_{\rm
m0}$--$\Omega_{k0}$ plane for the non-flat $\Lambda$CDM model is
shown in the lower-right panel. Interestingly, we find that, since
$c$ and $\Omega_{k0}$ are anti-correlated, and $\Omega_{\rm m0}$ and
$\Omega_{k0}$ are also anti-correlated, $c$ and $\Omega_{\rm m0}$ in
the $\Lambda(t)$CDM model have a positive correlation in this case.

\begin{figure*}[htbp]
\centering
\begin{center}
$\begin{array}{c@{\hspace{0.2in}}c} \multicolumn{1}{l}{\mbox{}} &
\multicolumn{1}{l}{\mbox{}} \\
\includegraphics[scale=0.25]{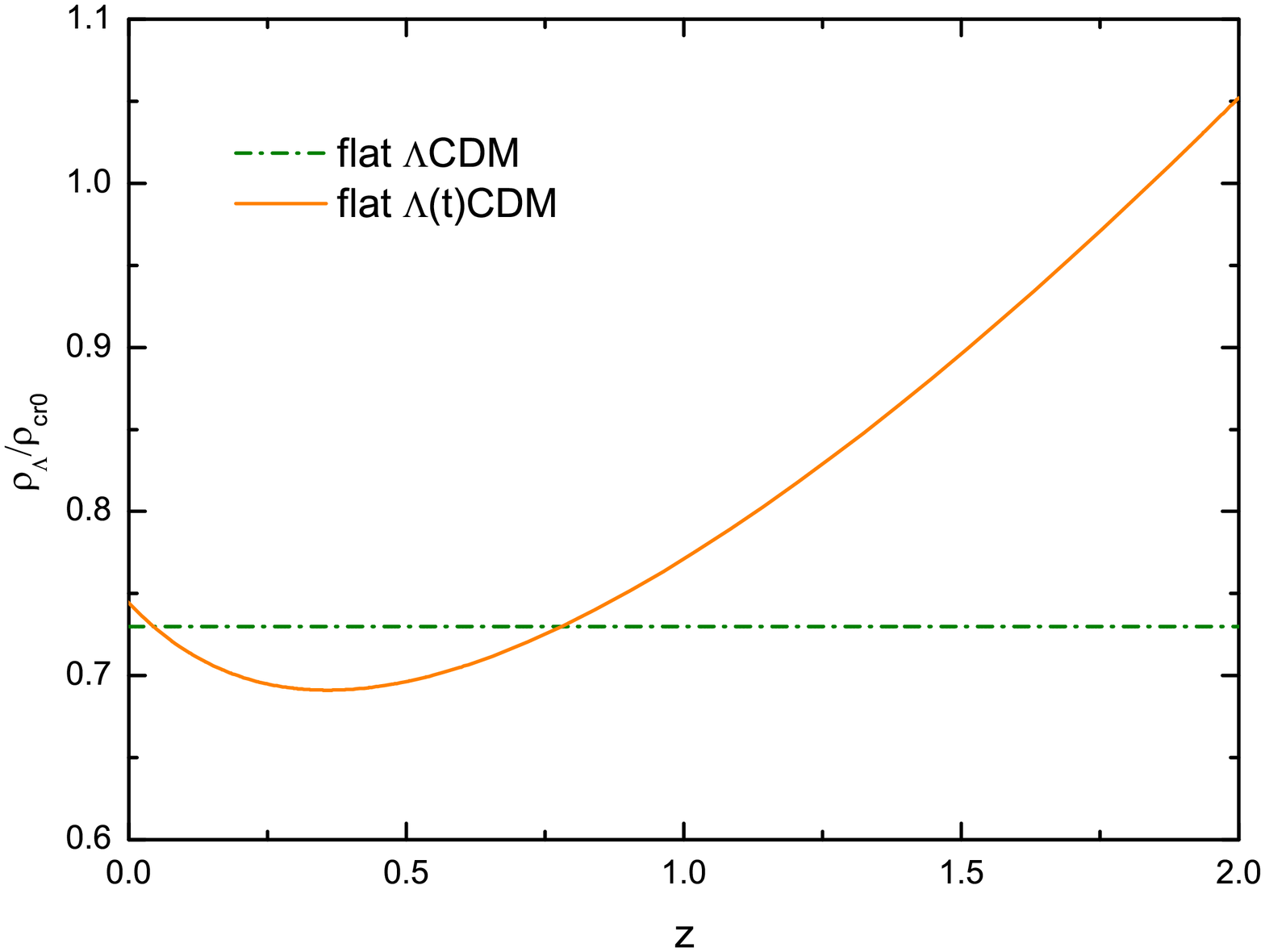} &\includegraphics[scale=0.25]{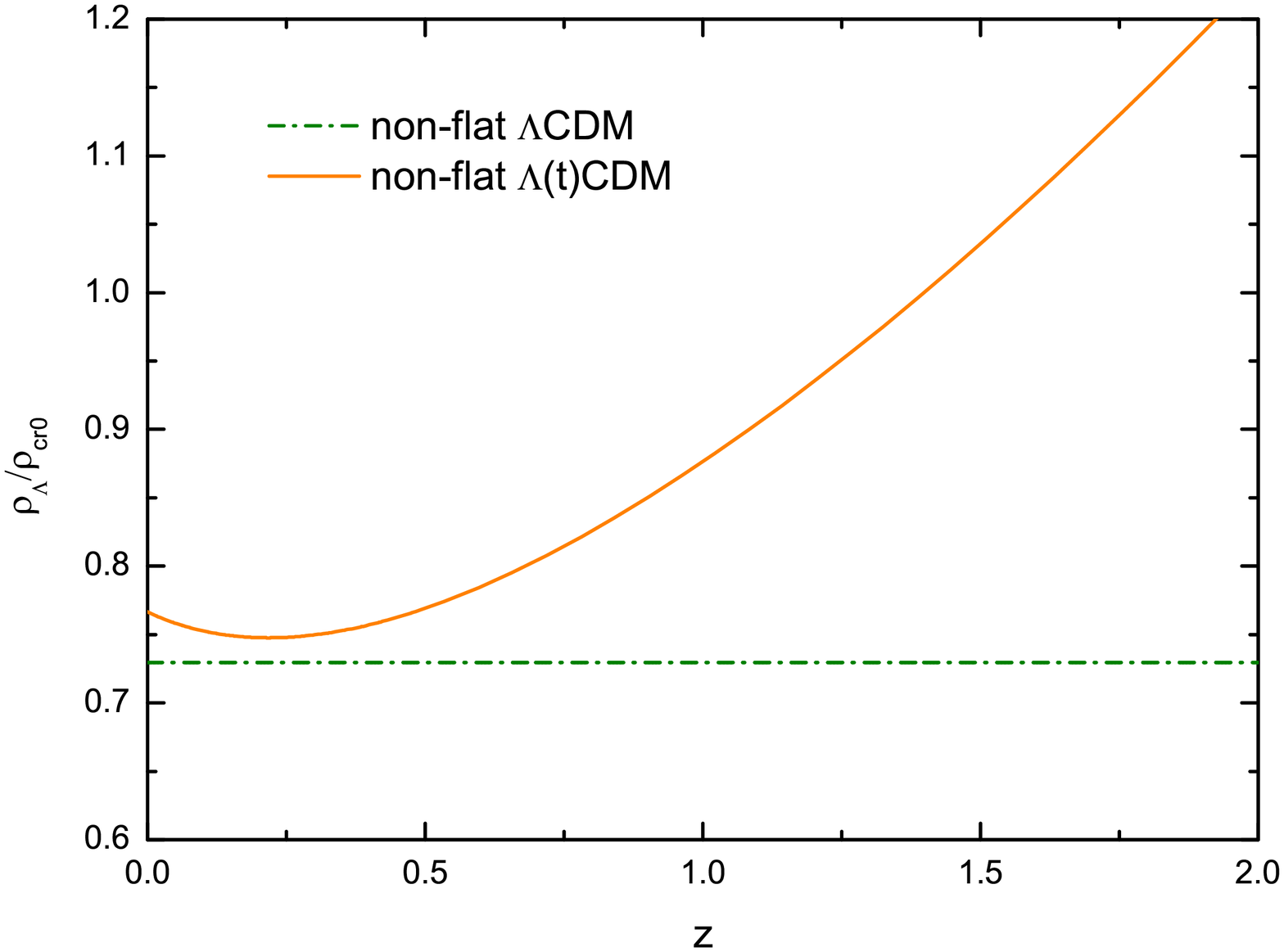} \\
\end{array}$
\end{center}
\caption[]{\small \label{fig3}The evolutions of
$\rho_\Lambda/\rho_{\rm cr0}$ in the holographic $\Lambda(t)$CDM and
$\Lambda$CDM models, in the best-fit cases.}
\end{figure*}

\begin{figure*}[htbp]
\centering
\begin{center}
$\begin{array}{c@{\hspace{0.2in}}c} \multicolumn{1}{l}{\mbox{}} &
\multicolumn{1}{l}{\mbox{}} \\
\includegraphics[scale=0.25]{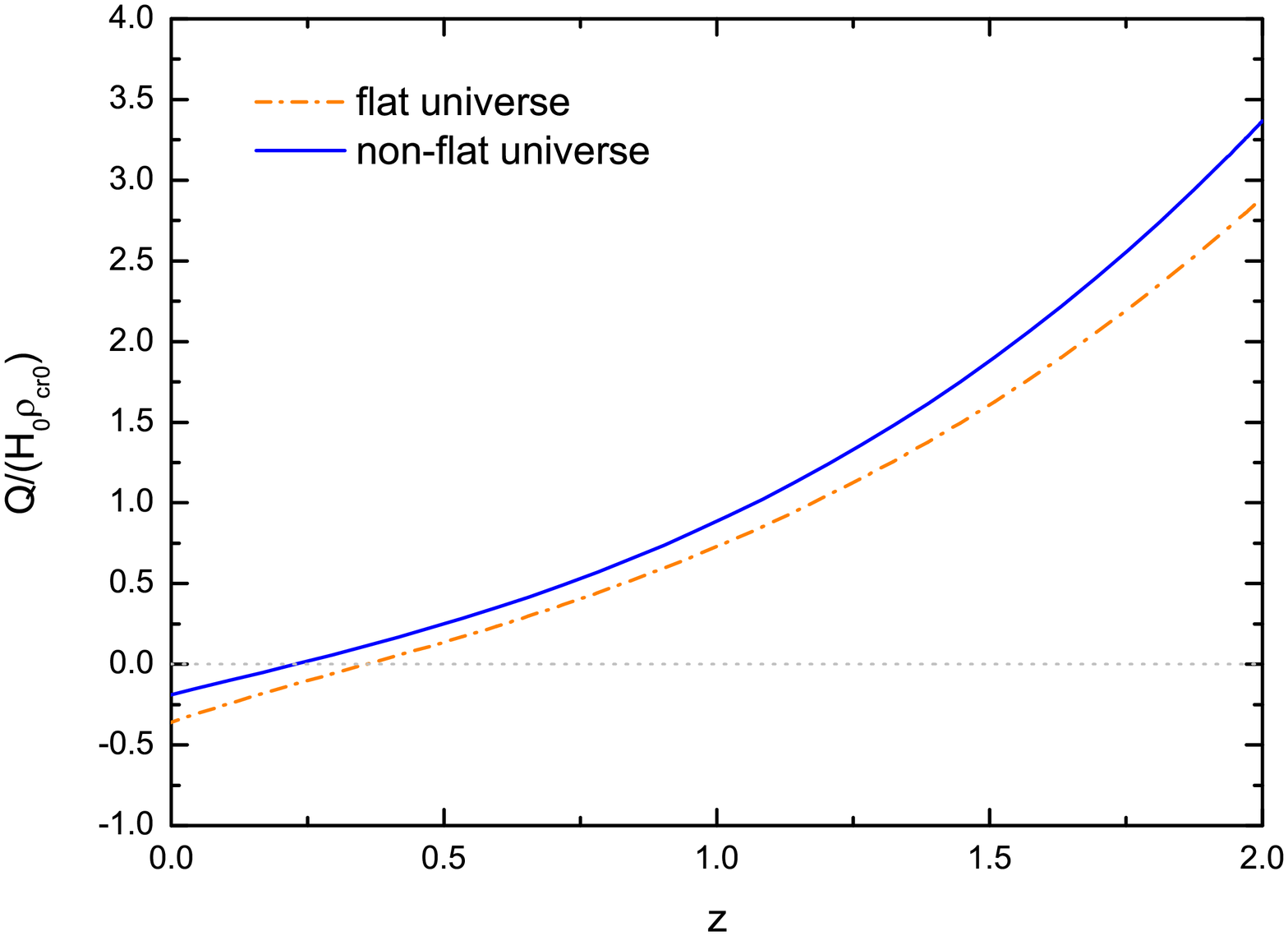} &\includegraphics[scale=0.25]{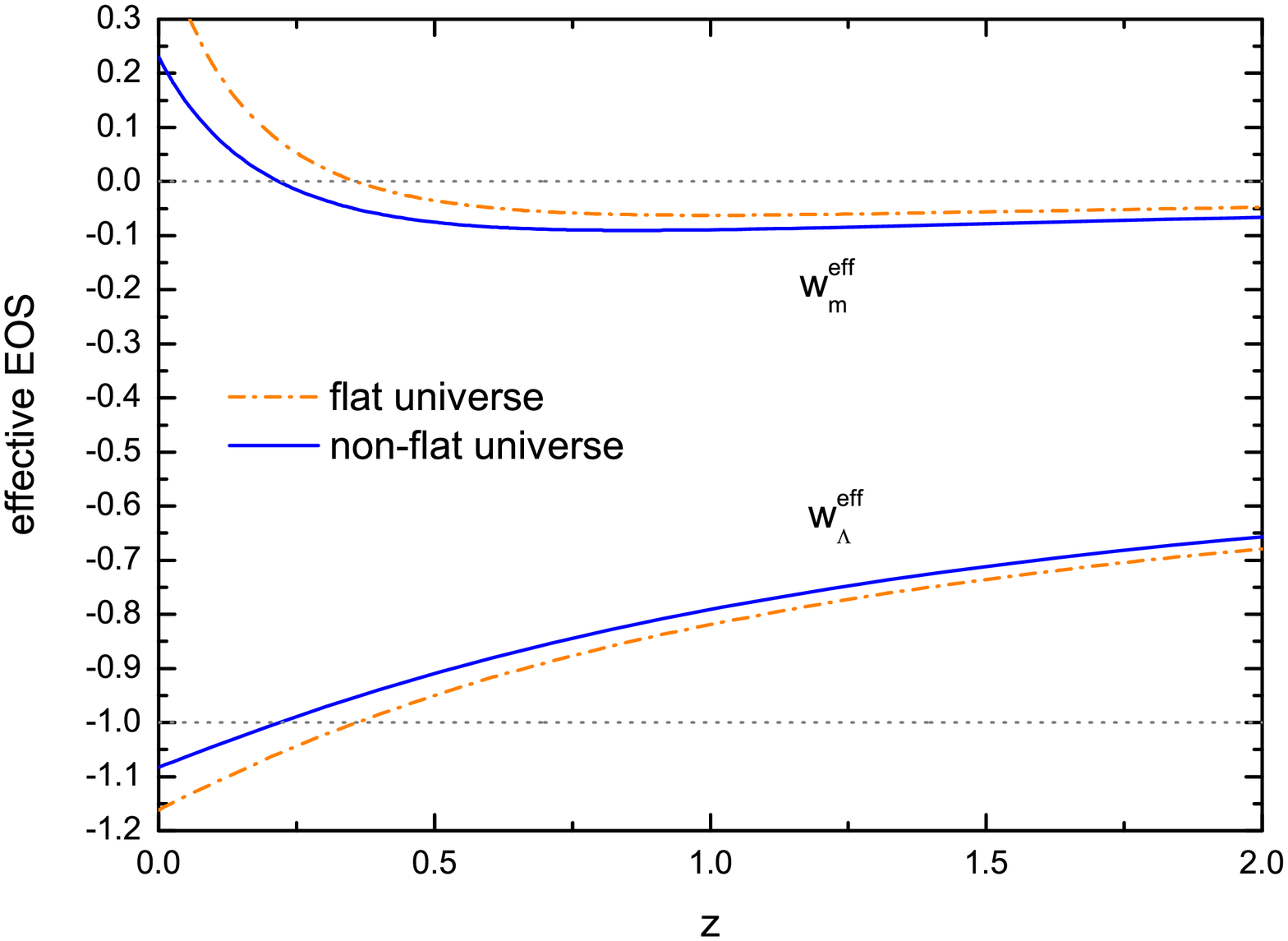} \\
\end{array}$
\end{center}
\caption[]{\small \label{fig4}The evolutions of $Q/(H_0\rho_{\rm
cr0})$ and the effective EOS parameters, $w_{\rm m}^{\rm eff}$ and
$w_{\Lambda}^{\rm eff}$, in the holographic $\Lambda(t)$CDM model,
in the best-fit cases.}
\end{figure*}

Next, we shall analyze the cosmological evolution of the holographic
$\Lambda(t)$CDM model using the fit results. In the standard
$\Lambda$CDM model, $\Lambda$ is a real constant, and so the vacuum
energy density $\rho_\Lambda$ remains constant during the expansion
of the universe. However, in the holographic $\Lambda(t)$CDM model,
though the EOS of the vacuum remains $-1$, the cold dark matter
component exchanges energy with the vacuum, so that the vacuum
energy density $\rho_\Lambda$ is not a constant but dynamically
evolves during the expansion of the universe. Thus, we want to see
how $\rho_\Lambda$ evolves in the $\Lambda(t)$CDM model. We want to
learn the direction of the energy flow between $\Lambda(t)$ and CDM.
For this purpose, we plot $\rho_\Lambda/\rho_{\rm cr0}$ versus $z$
in Fig.~\ref{fig3} by using the best-fit results, where $\rho_{\rm
cr0}=3M_{\rm Pl}^2H_0^2$ is the present-day critical density of the
universe. In general, we have $\rho_\Lambda/\rho_{\rm
cr0}=E^2\Omega_\Lambda$; for the $\Lambda$CDM model this quantity is
just $\Omega_{\Lambda 0}$, and for the holographic $\Lambda(t)$CDM
model this quantity can be directly obtained by using the solutions
of the differential equations (\ref{deq1}) and (\ref{deq2}). The
cases of the flat and non-flat universes are shown in the left and
right panels of Fig.~\ref{fig3}, respectively, where the solid lines
denote the $\Lambda(t)$CDM model and the horizontal dash-dot lines
correspond to the $\Lambda$CDM model. From this figure, we see that,
in the $\Lambda(t)$CDM model, $\rho_\Lambda$ is not a constant, but
is dynamical, first decreases and then increases during the
expansion of the universe, exhibiting a quintom feature. This
implies that the interaction between $\Lambda(t)$ and CDM induces an
energy flow of which the direction is first from $\Lambda(t)$ to CDM
and then from CDM to $\Lambda(t)$. To see the direction of the
energy flow more clearly, let us derive the explicit expression of
$Q$. Combining Eqs.~(\ref{coneq1}), (\ref{deqb1}), and
(\ref{deqb2}), we obtain
\begin{equation}
{Q\over H_0\rho_{\rm
cr0}}=-2E^3\Omega_\Lambda\left(\sqrt{{\Omega_\Lambda\over
c^2}+\Omega_k}-1\right).\label{Qterm}
\end{equation}
Then, we can directly plot $Q(z)$. Figure~\ref{fig4} shows the
evolutions of $Q/(H_0\rho_{\rm cr0})$ and the effective EOS
parameters, $w_{\rm m}^{\rm eff}$ and $w_{\Lambda}^{\rm eff}$, in
the $\Lambda(t)$CDM model, where the dash-dot lines denote the flat
universe and the solid lines correspond to the non-flat universe. We
see clearly that $Q$ crosses the noninteracting line ($Q=0$) from
$Q>0$ to $Q<0$. This indeed indicates that at first $\Lambda(t)$
decays to CDM, and then CDM decays to $\Lambda(t)$. We can also see
from the right panel of Fig.~\ref{fig4} that $w_{\rm m}^{\rm eff}$
crosses the $w=0$ line from $w<0$ to $w>0$, and $w_{\Lambda}^{\rm
eff}$ crosses the $w=-1$ line from $w>-1$  to $w<-1$, at around
$z=0.2-0.3$, conforming the sign-change of $Q$ discussed above. In
fact, in Ref.~\cite{Qsign}, it has been shown that the interaction
between dark energy and dark matter may change sign during the
cosmological evolution. By parameterizing the coupling and fitting
to the current data, it is shown in Ref.~\cite{Qsign} that a
time-varying vacuum scenario is favored, in which the interaction
$Q(z)$ crosses the noninteracting line during the cosmological
evolution and the sign changes from negative to positive. The
holographic $\Lambda(t)$CDM model is just a time-varying vacuum
energy model in which $Q$ changes sign, however, in this model the
sign changes from positive to negative, opposite to the case in
Ref.~\cite{Qsign}. This implies that we should pay more attention to
the time-varying vacuum model and seriously consider the theoretical
construction of a sign-changeable or oscillatory interaction between
dark sectors.

\section{Conclusion}\label{sec:concl}

In this paper, we have studied the holographic $\Lambda(t)$CDM model
in a non-flat universe. We demonstrated that, in the holographic
model, it is best to choose the event horizon size of the universe
as the IR cutoff of the effective quantum field theory. We then
derived the evolution equations of the model in a non-flat universe.
It is shown that the cosmological evolution of the model is governed
by the differential equations (\ref{deq1}) and (\ref{deq2}); solving
these two differential equations, one can obtain the evolutions of
$E(z)$ and $\Omega_\Lambda(z)$ directly, and other cosmological
quantities can subsequently be derived. Actually, the holographic
$\Lambda(t)$CDM model is an interacting holographic vacuum energy
scenario in which the interaction between $\Lambda(t)$ and CDM is
described by $Q=-\dot{\rho}_\Lambda$. The explicit expression of $Q$
is given by Eq.~(\ref{Qterm}).

In order to learn the cosmological evolution described by the
holographic $\Lambda(t)$CDM model, we place observational
constraints on the model, and analyze the cosmological evolution of
the model using the fit results. We constrained the holographic
$\Lambda(t)$CDM model and the $\Lambda$CDM model by using the
current observational data, including the 557 Union2 SN data, the
CMB WMAP 7-yr data, and the BAO SDSS data. Our fit results show that
the spatial curvature plays a significant role in the model in
fitting the data. Once the additional parameter $\Omega_{k0}$ is
involved in the model, the $\chi^2_{\rm min}$ value decreases by a
rather distinct amount, and correspondingly, the ranges of the
parameters amplify to some extent as usual. We found that the
holographic $\Lambda(t)$CDM model tends to favor a spatially closed
universe (the best-fit value of $\Omega_{k0}$ is $-0.042$), and we
found that the $95\%$ CL range for the spatial curvature is
$-0.101<\Omega_{k0}<0.040$. Moreover, we found that, for the
holographic $\Lambda(t)$CDM model in a flat universe, the parameters
$c$ and $\Omega_{\rm m0}$ are anti-correlated, however, when the
model is placed in a non-flat universe, $c$ and $\Omega_{\rm m0}$
get a positive correlation, in our data analysis.

We illustrated the cosmological evolution of the model by taking the
best-fit values of the parameters as an example. We showed that the
vacuum energy density $\rho_\Lambda$ is indeed not a constant in the
$\Lambda(t)$CDM model; it first decreases and then increases during
the expansion of the universe, exhibiting a quintom feature. This
implies that the interaction between $\Lambda(t)$ and CDM induces an
energy flow of which the direction is first from $\Lambda(t)$ to CDM
and then from CDM to $\Lambda(t)$. We also plotted the evolution of
$Q$ and showed that $Q$ crosses the noninteracting line ($Q=0$) from
$Q>0$ to $Q<0$, at around $z=0.2-0.3$. So, the holographic
$\Lambda(t)$CDM model is just a time-varying vacuum energy scenario
in which $Q$ changes sign.

\begin{acknowledgments}
This work was supported by the National Science Foundation of China
under Grant Nos.~10705041, 10975032, 11047112 and 11175042, and by
the National Ministry of Education of China under Grant
Nos.~NCET-09-0276, N100505001, N090305003, and N110405011.
\end{acknowledgments}


\end{document}